\documentstyle[preprint,pra,floats,aps,psfig]{revtex}

\begin{document}

\title{New access to very weak interactions in molecules}
\author{P.L.~Chapovsky\thanks{E-mail: chapovsky@iae.nsk.su}}

\address{  Institute of Automation and Electrometry,\\ 
           Russian Academy of Sciences, 630090 Novosibirsk, Russia}
         
\date{\today}

\draft
\maketitle

\begin{abstract}
It is predicted that nuclear spin conversion in molecules can be 
efficiently controlled by strong laser radiation resonant to 
rovibrational molecular transition. The phenomenon can be used for 
substantial enrichment of spin isomers, or for detection of very
weak ($10-100$~Hz) interactions in molecules.
\end{abstract}

\vspace{2cm}

\section{Introduction}

Many symmetrical molecules exist in Nature only in the form of nuclear spin
isomers \cite{Landau81}. These isomers differ by  symmetry of 
nuclear spin wave function and, consequently, by symmetry of 
molecular spatial wave 
function. Relaxation between different spin states (spin conversion) is
extremely slow process if it is compared with other gas kinetic rates,
e.g., vibrational relaxation. It makes nuclear spin isomers 
unique objects with many potential applications. 

It was shown in a recent paper \cite{Chap00PRA1} that external
radiation can influence conversion of spin isomers both through
level populations and through optically induced coherences. 
Purpose of the paper \cite{Chap00PRA1} was to investigate main features
of the coherent control of isomer enrichment and conversion.
In order to achieve the goal, the process was considered in a simplest 
arrangement in which microwave radiation excited molecular 
rotational transition.

In the present paper we will consider the process in more complicated 
arrangement which promises significantly better control 
of spin conversion. One outcome of this high efficiency
is that the coherent control in new arrangement can be used for 
detection of very weak ($10-100$~Hz) interactions in molecules.  

\section{Equation of change}

\begin{figure}[htb]
\centerline{\psfig
{figure=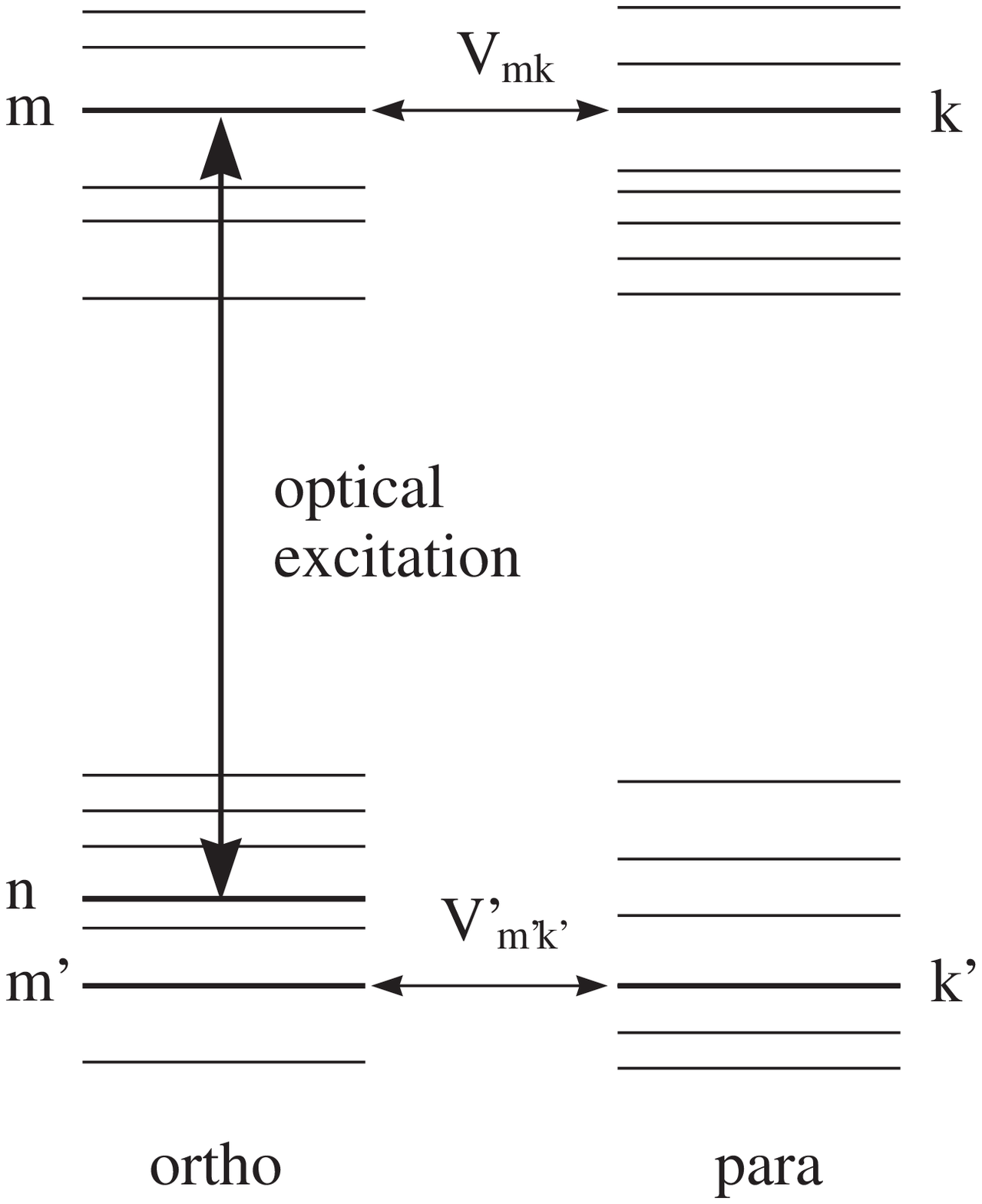,height=14cm}}
\vspace{2cm}
\caption{\sl Level scheme. Horizontal lines indicate the ortho-para mixing
in the ground and excited vibrational states. There are rotational relaxation
inside each vibrational state (rate $\nu_R$) and vibrational
relaxation from upper states (rate $\nu_V$).}
\label{fig1}
\end{figure}

First, we give qualitative picture of the process.
Let us assume that a test molecule has two nuclear spin states, ortho and
para, and  that there is a laser radiation
resonant to the rovibrational transition, $m-n$, in the ortho subspace,
Fig.~\ref{fig1}. The low state $n$ is not mixed with
para states, but the upper state $m$ is mixed  by the
{\it intramolecular} perturbation
$\hat V$ with the para state $k$. In addition, there is the ortho-para 
level pair, $m'-k'$,  in the ground vibrational state mixed by 
another {\it intramolecular} perturbation $\hat V'$. We assume 
that collisions of the test molecules cannot alter their spin state. 
This arrangement corresponds to the general formulation of 
{\it quantum relaxation} in which one has collisionally isolated
subspaces of states mixed together by internal perturbation
\cite{Chap96PA}. Suppose that the molecule is placed initially in 
the ortho subspace of the ground vibrational state.
Due to collisions the test molecule will undergo fast rotational
relaxation {\it inside} the ortho subspace. This will proceed until molecule 
jumps to the state $m'$, which is directly mixed with the para state $k'$, 
or to the state $n$ which is mixed by combined action of the 
external field and intramolecular perturbation
$\hat V$ with the para state $k$. Admixture of a para state 
implies that the next collision can move the molecule to another para 
states and thus localize it inside the para subspace.
It is clear that resonant radiation can significantly modify the spin 
conversion. One knows from the 
literature on resonant interaction of strong laser 
radiation with matter (see, e.g., \cite{Rautian79,Cohen-Tannoudji92}) 
that radiation can change population of states, split 
states and create coherences in the system. All these can
affect the mixing of ortho and para states and thus the spin
conversion process. 

Quantitative description of the problem can be performed with the help of 
kinetic equation for density matrix, $\hat\rho$.
The molecular Hamiltonian consists of four terms,
\begin{equation}
     \hat H = \hat H_0 + \hbar\hat G + \hbar\hat V + \hbar\hat V'.
\label{H}
\end{equation} 
The main part, $\hat H_0$, has the eigen ortho and para states 
shown in Fig.~\ref{fig1}. $\hbar\hat G$ describes 
the molecular interactions with the external radiation,
\begin{equation}
\hat G = - ({\bf E}_0\hat{\bf d}/\hbar)\cos\omega_Lt,   
\label{g}
\end{equation}
where ${\bf E}_0$ and $\omega_L$ are the amplitude and frequency 
of the electromagnetic wave; $\hat {\bf d}$ is the  
operator of the molecular electric dipole moment. We have neglected  
molecular motion in the operator $\hat G$ in order to simplify
the theory. 

In the representation of the eigen states of $\hat H_0$ kinetic equation reads,
\begin{equation}
     d\hat\rho/dt = (d\hat\rho/dt)_{coll} - i [\hat G+\hat V+\hat V',\hat\rho], 
\label{r1}
\end{equation}
where $(d\hat\rho/dt)_{coll}$ is the collision integral. Further, 
collisions in our system will be described by the model standard
in the theory of molecular interaction with laser radiation. The off-diagonal 
elements of $(d\hat\rho/dt)_{coll}$ will be assumed to have only decay terms,
\begin{equation}
     (d\rho_{j,j'}/dt)_{coll} = -\Gamma\rho_{jj'}; \ \ \  j\neq j'.
\label{S_off}
\end{equation}
Here, $j$ and $j'$ indicate rovibrational states of the molecule.
These states are assumed to have no degeneracy. The decoherence rates
were taken equal for all off-diagonal elements of collision integral.
Diagonal terms of the collision integral will be described in the framework of
the strong collision model.

Our goal is to determine time dependence of the total concentration
of molecules in one spin state. 
For example, for the total concentration of ortho molecules, $\rho_o$, 
one can get from Eq.~(\ref{r1}) the following equation of change, 
\begin{equation}
     d\rho_o/dt = 2Re\,i(\rho_{mk}V_{km}+\rho_{m'k'}V'_{k'm'}).
\label{ro}
\end{equation} 
In fact, this result is valid for any model of collision integral, as 
long as collisions do not alter the molecular spin state which implies, 
$\sum_j(d\rho_{jj}/dt)_{coll}=0$, if $j\in$~ortho, or $j\in$~para.

One has to make a few simplifications in order to find the off-diagonal 
density matrix elements, $\rho_{mk}$ and $\rho_{m'k'}$. We assume $\hat V$ and 
$\hat V'$ being small and consider zero and first order 
terms of the density matrix, 
\begin{equation}
     \hat\rho = \hat\rho^{(0)} + \hat\rho^{(1)}.
\label{split}
\end{equation} 
We start with zero order perturbation theory. $\rho^{(0)}$ is determined by the
kinetic equation,
\begin{equation}
       d\hat\rho^{(0)}/dt = (d\hat\rho^{(0)}/dt)_{coll} - i [\hat G,\hat\rho^{(0)}].
\label{ro0}
\end{equation}
In zero order perturbation theory, para molecules are at equilibrium,
\begin{equation}
     \rho^{(0)}_p(g,j) = (n-\rho^{(0)}_o) w(j);\ \ \ \rho^{(0)}_p(e,j) = 0, 
\label{0rp}
\end{equation}
where $n$ is the total concentration of the test molecules; $w(j)$ is
the Boltzmann distribution over rotational states. We will assume the
same function $w(j)$ for each of four vibrational states. We have 
neglected in Eq.~(\ref{0rp}) vibrational resonance exchange between 
ortho and para molecules which would populate the upper vibrational 
state of para molecules. It was taken into account also that ortho-para
exchange is on many orders of magnitude slower than vibrational and
rotational relaxations.

Equations for stationary level populations of ortho molecules are obtained 
from Eqs.~(\ref{S_off}) and (\ref{ro0}). Thus one has,
\begin{eqnarray}
(\nu_V + \nu_R)\rho^{(0)}_o(e,j) & = & \nu_R w(j)\rho^{(0)}_o(e)    
         + \rho^{(0)}_o p\,\delta_{jm}; \nonumber \\
          \nu_R\rho^{(0)}_o(g,j) & = &  \nu_R w(j)\rho^{(0)}_o(g) 
          + \nu_V w(j)\rho^{(0)}_o(e)  
          - \rho^{(0)}_o p\,\delta_{jn}; \nonumber \\
 \rho^{(0)}_o p & = & \frac{2\Gamma G^2}{\Gamma^2 + \Omega^2}
                \left[\rho^{(0)}_o(g,n)  - \rho^{(0)}_o(e,m)\right],      
\label{sys0}
\end{eqnarray}
where $\rho^{(0)}_o(e)$ and $\rho^{(0)}_o(g)$ are the concentrations
of ortho molecules in excited and ground vibrational states; 
$\nu_V$ and $\nu_R$ are the rates of vibrational and rotational relaxations;
$p$ is the probability of optical excitation of ortho molecules. 
Introduction of different relaxation rates for different degrees
of freedom makes the model of strong collisions more accurate.
Eqs.~(\ref{sys0}) correspond to rotational wave approximation. 
Matrix element of $\hat G$ is given by
\begin{equation}
     G_{mn} = - Ge^{-i\Omega t};\ \ 
     G\equiv E_0 \overline{d}_{mn}/2\hbar, 
\label{rwa}
\end{equation}
where $\Omega=\omega_L - \omega_{mn}$ is the radiation frequency
detuning from the absorption line center, $\omega_{mn}$; 
the line over symbol indicate a time-independent factor. 
Rabi frequency, $G$, is assumed to be real.

Solution of Eqs.~(\ref{sys0}) has no difficulty. Concentration of
ortho molecules in excited vibrational state, $\rho^{(0)}(e)$, and
in the state $m$  which one needs  for further calculations read,
\begin{eqnarray}
\rho^{(0)}_o(e) & = & \rho^{(0)}_o \frac{p}{\nu_V};\ \ \  
                \rho^{(0)}_o(e,m) = \rho^{(0)}_o\frac{p}{\nu_m};\ \ \ 
                 p = \frac{2\Gamma G^2 w(n)}{\Gamma^2_B+\Omega^2}; \nonumber \\  
 \nu^{-1}_m & = & w(m)\nu_V^{-1} + (1-w(m))(\nu_V+\nu_R)^{-1}.               
\label{sol0}
\end{eqnarray}
Here, $\Gamma_B$ is the homogeneous linewidth of the absorption 
spectrum profile, $\Gamma^2_B = \Gamma^2 + 2\Gamma\tau G^2;\  
\tau = \nu^{-1}_m + \nu^{-1}_n;\ 
\nu^{-1}_n = w(n)\nu_V^{-1} + (1-w(n))\nu_R^{-1}$.
$\nu_m$ and $\nu_n$ are the effective population decay rates
of the corresponding states. In a similar way, one can calculate from Eq.~(\ref{ro0})
the off-diagonal density matrix element, which amplitude is equal to,
\begin{equation}
     \overline\rho^{(0)}_o(m|n) = i G \rho^{(0)}_o w(n)
     \frac{\Gamma+i\Omega}{\Gamma^2_B+\Omega^2}.
\label{ro_off}
\end{equation} 

In zero order perturbation theory, one neglects perturbations
$\hat V$ and $\hat V'$. It implies that there are no coherences
between ortho and para states, $\rho^{(0)}_{mk}=0$;  
$\rho^{(0)}_{m'k'}=0$. Consequently, one has,
\begin{equation}
     d\rho_o/dt = 2Re\, i(\rho^{(1)}_{mk}V_{km}+\rho^{(1)}_{m'k'}V'_{k'm'}),
\label{ro2}
\end{equation} 
instead of Eq.~(\ref{ro}). Note, that the spin conversion appears
in the second order approximation. The first order correction to the 
density matrix, $\rho^{(1)}$, is determined by the equation, 
\begin{equation}
     d\hat\rho^{(1)}/dt  = (d\hat\rho^{(1)}/dt)_{coll}
        - i [\hat G,\hat\rho^{(1)}] - i [\hat V + \hat V',\hat\rho^{(0)}].
\label{ro1}
\end{equation}
For $\rho^{(1)}_{m'k'}$ one has from this equation,
\begin{equation}
     \rho^{(1)}_{m'k'} = \frac{-iV'_{m'k'}}{\Gamma + i\omega'}
     [\rho^{(0)}_p(g,k')-\rho^{(0)}_o(g,m')],  
\label{1rho}
\end{equation}
where $\omega'\equiv\omega_{m'k'}$.
$\rho^{(1)}_{mk}$ can be obtained from equations
which are deduced from Eq.~(\ref{ro1}),
\begin{eqnarray}
d\rho^{(1)}_{mk}/dt + \Gamma\rho^{(1)}_{mk} + iG_{mn}\rho^{(1)}_{nk} & = & 
            iV_{mk} \rho^{(0)}_o(m); \nonumber \\
d\rho^{(1)}_{nk}/dt  + \Gamma\rho^{(1)}_{nk} + iG_{nm}\rho^{(1)}_{mk}  & = &  
             iV_{mk} \rho^{(0)}_o(n|m).    
\label{sys1}
\end{eqnarray}
Substitutions, 
     $V_{mk}=\overline{V}e^{i\omega t},\ \ (\omega\equiv\omega_{mk});\ \ 
     \rho^{(1)}_{mk} = \overline{\rho}^{(1)}_{mk}e^{i\omega t};\ \ 
     \rho^{(1)}_{nk} = \overline{\rho}^{(1)}_{nk}e^{i(\omega_L-\omega_{kn} t)}$,
transform Eqs.~(\ref{sys1}) to algebraic equations from which one 
finds $\rho^{(1)}_{mk}$. Using $\rho^{(1)}_{mk}$ and $\rho^{(1)}_{m'k'}$ 
from Eq.~(\ref{1rho}) one has an equation of change (\ref{ro2}) in the form,
\begin{eqnarray}
 \frac{d\rho_o}{dt} & = & \frac{2\Gamma|V'|^2}{\Gamma^2+\omega'^2}
                \left[\rho^{(0)}_p(g,k')-\rho^{(0)}_o(g,m') \right] - \nonumber \\ 
            &&2|V|^2 Re\frac{[\Gamma+i(\Omega+\omega)]\rho^{(0)}_o(m) +
     iG\overline\rho^{(0)}_o(n|m)}{(\Gamma+i\omega)[\Gamma+i(\Omega+\omega)]+G^2}.    
\label{ro3}
\end{eqnarray}

\section{Enrichment and conversion}

\begin{figure}[htb]
\centerline{\psfig
{figure=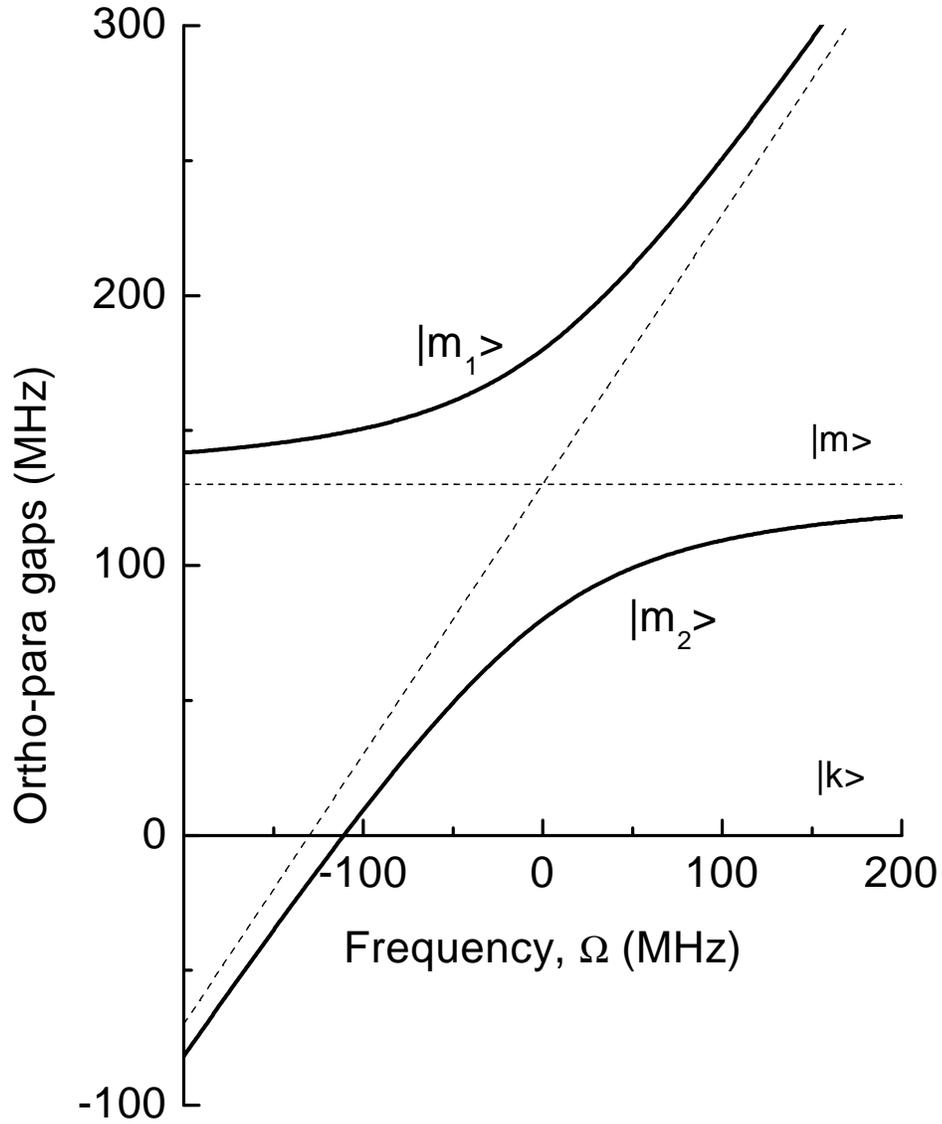,height=20cm}}
\vspace{2cm}
\caption{\sl Gaps between the ortho states $|m_1>$ and $|m_2>$
and para state $|k>$. Rabi frequency was taken equal $G=50$~MHz
and the radiation free ortho-para gap, $\omega=130$~MHz.}
\label{fig2}
\end{figure}

The denominator of the second term in the right-hand side of 
Eq.~(\ref{ro3}) is convenient to present as, $(\Gamma+i\omega_1)
(\Gamma+i\omega_2)$, where
\begin{equation}
     \omega_{1,2}=\omega + \frac{\Omega}{2}\pm
     \sqrt{\left(\frac{\Omega}{2}\right)^2+G^2}.
\label{denom}
\end{equation}
New parameters, $\omega_1$ and $\omega_2$, can be interpreted as the gaps
between the two components, $|m_1>$ and $|m_2>$, of the ortho state $|m>$, 
split by the optical field, and the para state $|k>$. The ortho state, 
$|m_2>$, crosses the para state, $|k>$, at $\Omega=-\omega(1-G^2/\omega^2)$,
see Fig.~\ref{fig2}. The splitting of states by resonant laser radiation
is well-known phenomenon in nonlinear spectroscopy
\cite{Rautian79,Cohen-Tannoudji92}. 

Using Eq.~(\ref{ro3}) one can present equation of change in the final form,
\begin{equation}
     d\rho_o/dt  =  n\gamma'_{op} - \rho_o\gamma;\ \ \ 
     \gamma  \equiv  \gamma'_{op} + \gamma'_{po} - \gamma'_n + \gamma_n + \gamma_{coh}.
\label{final}
\end{equation}
In writing this equation we have neglected in the right-hand side of Eq.~(\ref{ro3}) 
small difference between $\rho^{(0)}_o$ and the total
concentration of ortho molecules, $\rho_o$. In Eq.~(\ref{final}) the following 
partial conversion rates have been introduced. The field independent 
rates, 
\begin{equation}
     \gamma'_{op}=\frac{2\Gamma|V'|^2}{\Gamma^2+\omega'^2}w(k');\ \ \ 
     \gamma'_{po}=\frac{2\Gamma|V'|^2}{\Gamma^2+\omega'^2}w(m').
\label{g'}
\end{equation}
The rate $\gamma_{free}\equiv \gamma'_{op} + \gamma'_{po}$ determines 
the equilibration rate in the system without an external field. The field 
dependent term, 
\begin{equation}
     \gamma'_n = \gamma'_{po}p/\nu_V,
\label{g'n}
\end{equation}
appears because of depletion of the ground vibrational state 
of ortho molecules by optical excitation, Index $n$ in $\gamma'_n$ 
comes from ``noncoherent'', i.e., induced by 
level populations. Another term of similar ``noncoherent'' origin 
appears due to the level population, $\rho^{(0)}_o(m)$,  in Eq.~(\ref{ro3}),
\begin{equation}
     \gamma_n = 2|V|^2\frac{p}{\nu_m}Re\frac{\Gamma+i(\Omega+\omega)}
                {(\Gamma+i\omega_1)(\Gamma+i\omega_2)}.
\label{gn}
\end{equation}
And finally the ``coherent'' term, $\gamma_{coh}$, originated from
$\overline\rho^{(0)}_o(n|m)$, in Eq.~(\ref{ro3}),
\begin{equation}
     \gamma_{coh} = 2|V|^2\frac{p}{2\Gamma}Re\frac{\Gamma-i\Omega}
                {(\Gamma+i\omega_1)(\Gamma+i\omega_2)}.
\label{gcoh}
\end{equation}

Solution to Eq.~(\ref{final}) can be presented as,
\begin{equation}
    \rho_o = \overline\rho_o + (\rho_o(0)-\overline\rho_o)\exp(-\gamma t);\ \ \  
    \overline\rho_o = n\gamma'_{op}/\gamma.
\label{overro}
\end{equation}
Here $\gamma$ is the equilibration rate in the system
in the presence of external field; $\overline\rho_o$
is the stationary concentration of ortho molecules.
Without an external radiation (at the instant $t=0$), the equilibrium
concentration of para molecules is equal to,
\begin{equation}
     \rho_p(0)=n-\rho_o(0)=n\gamma'_{po}/\gamma_{free},
\label{rop}
\end{equation}   
if the Boltzmann factors are assumed to be equal, $w(k')=w(m')$. 
This implies $\gamma'_{op}=\gamma'_{po}$ (see Eq.~(\ref{g'})), 
the laser field produces a stationary enrichment of para molecules,
\begin{equation}
     \beta \equiv \frac{\overline\rho_p}{\rho_p(0)}-1 = 
     1-2\frac{\gamma'_{op}}{\gamma}.
\label{bp}
\end{equation}
One can see from this equation that external field changes concentration
of para isomers if $\gamma \neq \gamma_{free}$.

We assume in further analysis the following parameters,
$\omega=100$~MHz, $\omega'=130$~MHz, $V'_{m'k'}=5$~kHz, 
$\Gamma=2\cdot10^8$~s$^{-1}$/Torr and the 
Boltzmann factors of the states $m'$, $k'$, $m$, and $k$ all equal
$10^{-2}$. This set of parameters gives the field free conversion rate, 
$\gamma_{free}=10^{-2}$~s$^{-1}$/Torr, which coincides with the  
conversion rate in $^{13}$CH$_3$F. Nuclear spin conversion in these
molecules is governed by quantum relaxation (see the review \cite{Chap99ARPC}). 
The rotational and vibrational relaxation rates will be taken equal, 
$\nu_R=0.1\Gamma$ and $\nu_V=0.01\Gamma$, respectively.

\begin{figure}[htb]
\centerline{\psfig
{figure=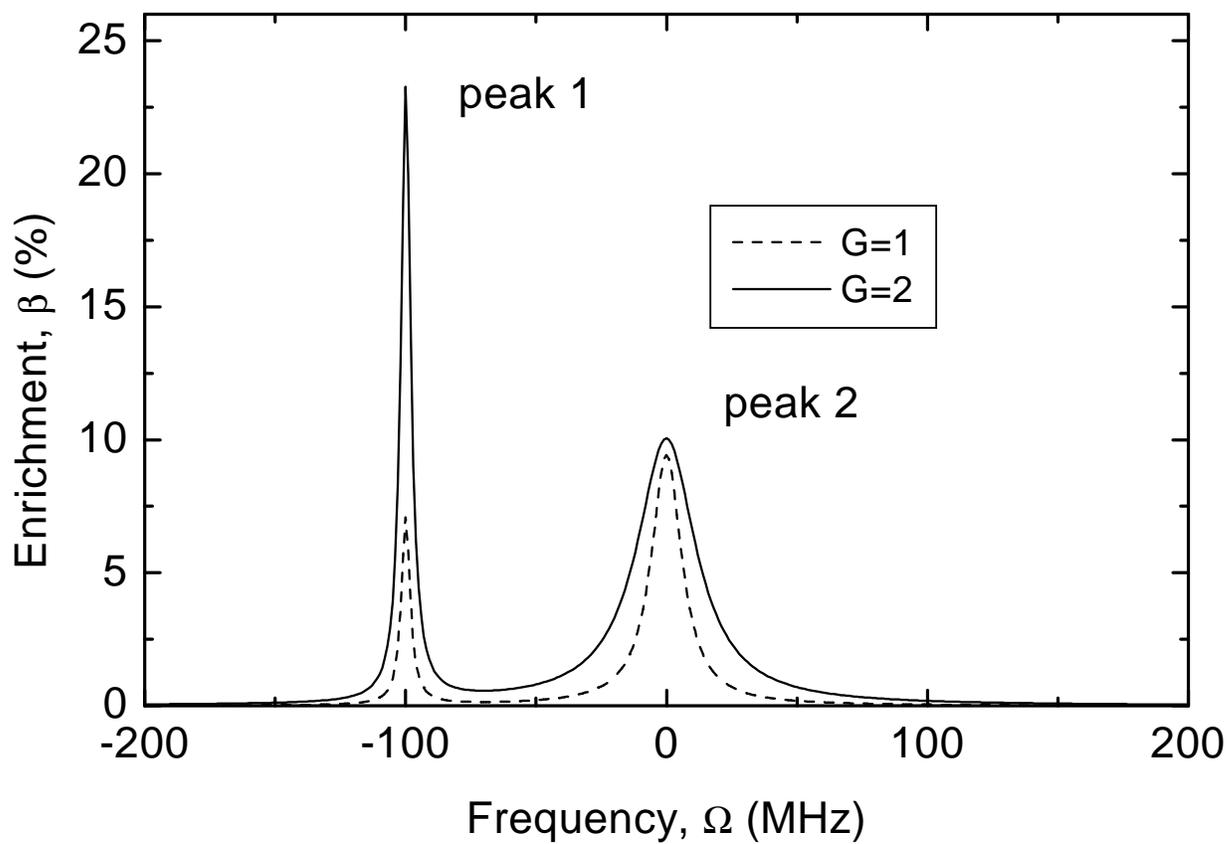,height=14cm}}
\vspace{2cm}
\caption{\sl Frequency dependence of the enrichment of para molecules, 
$\beta$, at $G=1$~and~2~MHz.}
\label{fig3}
\end{figure}

First, we consider relatively low optical fields, thus small $G$.
In this case one has two peaks in enrichment at frequencies $\Omega\simeq-\omega$
and $\Omega=0$, see Fig.~\ref{fig3}. The data shown in this figure
correspond to $V_{mk}=3$~kHz, and $\Gamma=2$~MHz. The peak at $\Omega=0$
appears because the excitation probability, $p$, has maximum at
this frequency. Amplitude of this peak is determined mainly by the rate $\gamma_n$. 
As $G$ grows, the amplitude of the peak~2 reaches the value $\gamma_n/\gamma_{free}
\sim (V\omega'/2V'\omega)^2$ which constitutes $\simeq 15\%$. 
The peak at $\Omega=0$ in isomer enrichment was predicted in
\cite{Ilichov98CPL} by considering only the level population effects. 

Peak at $\Omega\simeq -\omega$ appears because the ortho state $|m_2>$ 
crosses  the para state $|k>$ at this frequency of the external field
(see Fig.~\ref{fig2}). This peak is determined mainly by
$\gamma_{coh}$. When $G$ increases its amplitude grows to much bigger
values than the amplitude of the peak at $\Omega=0$. 
At resonant frequency $\Omega\simeq -\omega$ the rate
$\gamma_{coh}$ is enhanced by large factor $(\omega/\Gamma)^2$.
This explains much larger enrichment at  $\Omega\simeq -\omega$.
Note, that large enrichment occurs only if the excitation probability
at this frequency, $p(-\omega)$, is not very low.

\begin{figure}[htb]
\centerline{\psfig
{figure=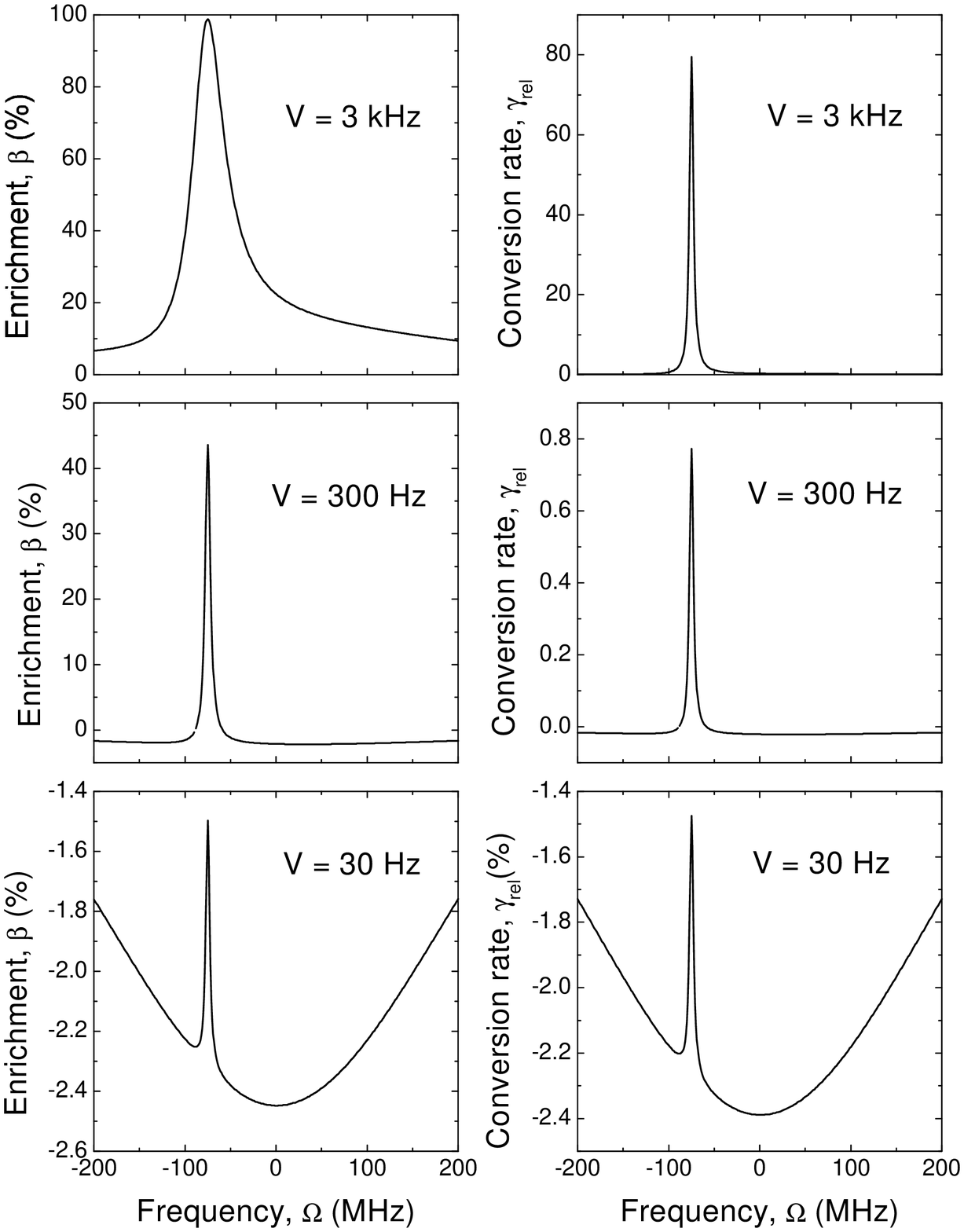,height=20cm}}
\vspace{2cm}
\caption{\sl Enrichment of para molecules, $\beta$, and conversion rate, 
$\gamma_{rel}$, for three values of $V_{mk}$.}
\label{fig4}
\end{figure}

The data shown in Fig.~\ref{fig4} correspond to strong optical field, 
$G=50$~MHz, and three values of $V_{mk}$. $\Gamma$ was taken equal 2~MHz. 
One can see, that strong optical field is able to convert
almost all molecules to the para state if $V_{mk}\simeq V_{m'k'}$.
Thus relatively weak (3~kHz) coupling in upper state is able to produce 
macroscopic effect, viz., almost complete enrichment of spin isomers.
It is of fundamental importance, that even for much weaker coupling
in upper state, enrichment is still significant. For example, if the
perturbation in upper state, $V_{mk}=30$~Hz, one has the 
enrichment, $\beta\simeq1$\%. Enrichment at this level can 
easily be measured.  It is important that the enrichment peak at 
$\Omega\simeq -\omega$ is narrow (the width $\simeq \Gamma$) 
and thus can be distinguished from much wider structures
(the width $\simeq \Gamma_B$) induced by population effects.

Equilibration rate in the system is given by $\gamma$, see Eq.~(\ref{overro}). 
It is convenient to characterize the conversion rate in relative units,
\begin{equation}
     \gamma_{rel} = \gamma/\gamma_{free} - 1,
\label{grel}
\end{equation}
Conversion rate, like enrichment, has two peaks in its frequency
dependence at low $G$. If Rabi frequency, $G$, is large and the ortho-para couplings
in upper and low states have the same order of magnitude, conversion can be
significantly enhanced (Fig.~\ref{fig4}, upper panel). Again, this 
enhancement appears because of the crossing of ortho and para states in
upper vibrational state by external field.

\section{Discussion}

The phenomenon considered in the paper is based on the level splitting
produced by resonant electromagnetic radiation. Sometimes, this splitting 
is called in optics the dynamic Stark effect. The essence of the effect 
can be understood as follows.  Mixing  of ortho and para states 
depends on magnitude of the perturbation $\hat V$ but also on the 
ortho-para level gap. Optical field splits the molecular state and thus
change the gaps between the ortho and para states. Conversion rate is
significantly enhanced when the ortho and para states cross. Similar 
enhancement occurs when ortho and para states are crossed
by ordinary Stark effect in an external DC electric field \cite{Nagels96PRL}.

In the same way, one can understand high
sensitivity of the phenomenon to weak ortho-para couplings
in excited state. Equilibrium concentrations of ortho and para molecules 
are achieved when one has the ortho-to-para flux in excited 
state equal to the back flux in the ground vibrational state. Back flux 
is slow because it is determined by the non-degenerate ortho-para level pair. 
On the other hand, the flux in excited state can be significantly enhanced by 
proper choice of radiation parameters which allows 
to cross the ortho and para states in upper vibrational state.

High efficiency of the proposed enrichment method can be used to 
detect weak perturbations in excited vibrational state. First of all,
it can be the hyperfine perturbations of the same origin and similar
magnitude as the perturbations in the ground vibrational state. 
We have seen that hyperfine coupling of the order of $\sim10^3$~Hz
is able to convert almost all molecules in one spin state. One can also  
detect much weaker interactions in molecules. In this case
one should select the ortho-para level pair in excited vibrational state  
which is not mixed by ``ordinary'' hyperfine interactions 
in order to avoid the weak interaction to be hidden by stronger, ordinary
hyperfine interactions. An interesting case is the crossings of states
having opposite parity. Mixing of such ortho and para states can be performed only
by spin-dependent, parity-odd interactions which are not observed 
in molecules yet.  

\section{Conclusions}

We have performed analysis of the spin isomer enrichment
and conversion governed by molecular 
rovibrational excitation. This analysis was done using
a few simplifications. We have neglected the Doppler broadening
of the absorbing transition, degeneracy of molecular states,
and resonant vibrational exchange between excited and unexcited molecules.
These simplifications are not crucial for the existence of the phenomenon. 
More detailed analysis will be done elsewhere.

We have shown that coherent control of nuclear spin conversion in
molecules can be efficiently performed by strong radiation
resonant to rovibrational
molecular transition. A possible applications of this phenomenon 
is the enrichment of molecular spin isomers. Another application is the
detection of very weak ($10-100$~Hz) interactions in molecules, which 
can be, e.g., parity-odd interactions.

\section*{Acknowledgments}

This work was made possible by financial support from the  
Russian Foundation for Basic Research (RFBR), grant No. 98-03-33124a.

\newpage

\newpage

\newpage

\newpage


\begin{thebibliography}{1}

\bibitem{Landau81}
L.~D. Landau and E.~M. Lifshitz, {\em Quantum Mechanics, 3rd ed.} (Pergamon
  Press, Oxford, 1981).

\bibitem{Chap00PRA1}
P.~L. Chapovsky, http://arXiv.org/abs/physics/0011012  .

\bibitem{Chap96PA}
P.~L. Chapovsky, Physica A (Amsterdam) {\bf 233},  441  (1996).

\bibitem{Rautian79}
S.~G. Rautian, G.~I. Smirnov, and A.~M. Shalagin, {\em Nonlinear resonances in
  atom and molecular spectra} (Nauka, Siberian Branch, Novosibirsk, Russia,
  1979), p.\ 310.

\bibitem{Cohen-Tannoudji92}
C. Cohen-Tannoudji, J. Dupont-Roc, and G. Grynberg, {\em Atom-Photon
  Interactions} (Wiley, New-York, 1992).

\bibitem{Chap99ARPC}
P.~L. Chapovsky and L.~J.~F. Hermans, Annu. Rev. Phys. Chem. {\bf 50},  315
  (1999).

\bibitem{Ilichov98CPL}
L.~V. Il'ichov, L.~J.~F. Hermans, A.~M. Shalagin, and P.~L. Chapovsky, Chem.
  Phys. Lett. {\bf 297},  439  (1998).

\bibitem{Nagels96PRL}
B. Nagels, N. Calas, D.~A. Roozemond, L.~J.~F. Hermans, and P.~L. Chapovsky,
  Phys. Rev. Lett. {\bf 77},  4732  (1996).

\end{thebibliography}
\end{document}